# 浅谈实体建模：历史、现状与未来

邹强

浙江大学 CAD&CG 国家重点实验室，浙江省 杭州市 310027

**摘　　要**：实体建模技术组成了 CAD 软件的"功能心脏"，相关基础理论和算法是 CAD 发展历史上最关键的成果之一，成功回答了为使计算机能够辅助产品设计与制造，需要在计算机中存什么几何信息以及怎么存的问题。本文对实体建模的主要历史发展脉络做了简要介绍，同时对各发展阶段的关键问题以及其研究现状进行了讨论，最后选取了三个方向对实体建模的未来做出展望，重点关注从 Computer-Aided Design 到 Computer-Automated Design 的发展趋势。

**关　键　词**：CAD；几何建模；实体建模；参数化建模；直接建模；复杂曲面、结构建模；智能 CAD

## 1　前言

实体建模技术是 CAD（Computer-Aided Design）领域 60 余年发展过程中最重要的成果之一，其回答了为使计算机能够辅助产品设计制造，需在计算机中存什么几何信息和怎么存的问题。这是所有 CAX 任务能够进行的前提。

当前我国所面临的工业软件"卡脖子"问题中，作为 CAX 软件"功能心脏"的几何建模内核如何做到自主可控，弥补与国际主流内核如 Parasolid、ACIS 的巨大差距，是重中之重。而所谓几何建模内核实质上主要指实体建模算法库。为此，实体建模技术对我国具有重要意义。

本文将梳理实体建模的历史发展脉络，叙述其基础理论、关键算法与难点、以及这些难点的解决现状。同时，本文还将对实体建模的未来作出展望，重点关注 Computer-Aided Design 向 Computer-Automated Design 发展的趋势。

### 1.1　从 CAD 到几何建模

产品即人造物理实体，产品模型指这一实体的计算机表示（即一种数据结构），而 CAD 即是使用计算机来构建、查询和编辑产品模型。CAD 于上世纪 50 年代末在 MIT 被提出[1]，当时主要是为了满足两大需求：

(1) 构建产品的计算机模型并对之进行处理，以满足二战后发展起来的数控机床对自动生成加工路径[1]的需求[2]，这实质上是为了满足机器与机器间的协作需求；
(2) 构建一个可以让人和计算机一起协作进行工程设计的系统，其中人负责创造性任务，而计算机负责机械式任务[3,4]，如图 1 所示。这实质上是为了满足人与机器间的协作需求。

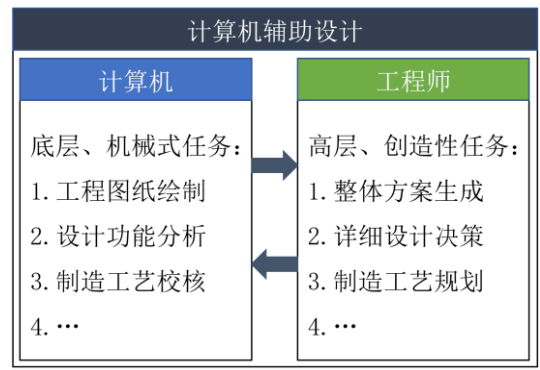

图 1　CAD 系统中人与计算机的分工

从这两个需求（尤其是第二个）出发，CAD 的具体内涵会非常丰富，包罗万象。但是不久后，人们意识到 CAD 应当聚焦于产品建模、产品分析和产品制造这三大内容，因为它们比其它方面更基础、本质[3]。由于同时期也有其它领域在研究产品分析和产品制造，最终这三大内容分流成我们今天熟

---

[1] 加工路径指机床刀具运动所遵循的轨迹[143–145]。

知的 CAD, CAE (Computer-Aided Engineering),
和 CAM (Computer-Aided Manufacturing)。

具体到 CAD，产品建模需要在计算机中构建能
够支撑产品全生命周期所需信息的数字模型。这些
信息以产品的几何形状为核心（如图 2 所示），并伴
有材料、工艺等非几何信息[5]。在实际计算机模型
中，材料、工艺等信息均可在几何模型的基础之上
以标记的形式来存储。正因为如此，CAD 建模往往
重点关注几何建模。到目前为止，主要的 CAD 几何
建模方法有：线框建模、曲面建模、实体建模、参
数化建模以及直接建模，分别简述如下。

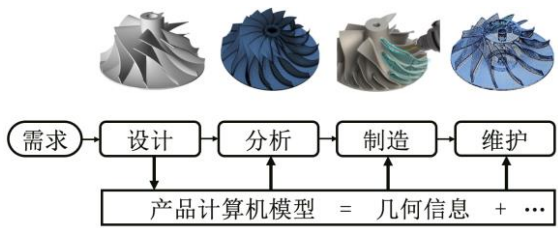

图 2  产品几何信息在产品全生命周期中的作用

## 1.2  从几何建模到实体建模

在众多几何建模方法[6]里面，线框建模到曲
面建模再到实体建模是 CAD 建模技术发展的初期阶
段，它们主要关注产品几何信息该以何种数学模型
来表示的问题。它们具体的发展脉络可以参见[7]
中图 3。

其中，线框模型，不管是 2D 还是 3D，利用产品
的边或轮廓来描述几何形状。2D 线框模型直接复制
传统工程制图，这也是第一代 CAD 原型系统
"Sketchpad"所采用的方式[8]。3D 线框模型的提
出是为了解决 2D 线框模型每变换一个视角，都要
重新绘制的问题（通过对 3D 线框模型进行投影操
作，可自动得到任意方向的视图）。然而，线框模型
存在两个重要缺陷：歧义性和无效性，如图 3 所示。
**无歧义性和确保有效性对 CAD 建模至关重要，因为
CAD 追求模型的真实性，这与追求真实感的图形学
不同[6,9]。**

为解决上述问题，人们提出曲面模型，对线框
模型进行"蒙皮"[10]。同时，二战后工业界对例
如汽车和飞机等复杂曲面设计制造的需求也促进
了曲面模型的发展[6]。人们提出了一系列巧妙的
曲面表示和操作方法，从 Coons 曲面到 Bezier 曲
面到 B-spline 曲面再到其改进型 NURBS 曲面，详
细的发展脉络见[11]。

然而，仅有曲面信息仍无法彻底解决歧义性和
无效性问题，例如真实世界中并不存在零厚度物体，
而且每个物体都有内外之分。为此，实体模型被提
出[12,13]，其特点在于对产品几何信息进行了完
整的表示，从点到边到面再到体。因其具有信息完
整性，产品是被唯一刻画的，同时任何几何性质
（如转动惯量）也都可以被自动计算[14]。

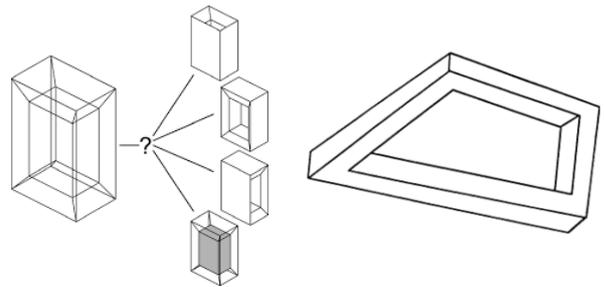

图 3  线框模型的缺陷：歧义性（左）和无效性
（右）[7]

## 1.3  从实体建模到参数化/直接建模

实体模型因其信息完整性而适合表示产品几
何信息。然而基本的实体建模方法，即 CSG
(Constructive Solid Geometry)[15] 和 B-rep
(Boundary Representation)[16]，使得实体模型一
旦被构建便难以修改[17]。因此，早期实体建模方
法一般仅被用作记录已经设计好并且不会发生变
动的产品，对整个设计过程，尤其是早期概念设计
阶段，帮助不大[7,18]。

为解决这一问题，参数化建模在 1980 年代末
被提出[2]。其基本思路是在实体模型的基础上添
加一层关联（associativity）信息，即在组成实体
模型的几何元素之间添加关联信息，如此，模型
上的局部变动可以按设计好的方式自动传播到模型
的其它区域[18]。关联信息一般以几何约束（例如
距离、相切、共轴等）的方式给出。这使得模型形
状被参数化到某几个控制参数上，即模型形状是这
些控制参数的一个函数[19,20]。这样，通过在实体
建模之上添加一层关联信息，人们获得了参数驱动
的实体模型变动能力。

值得一提的是，人们又在"实体+关联"的基础
之上，添加一层语义信息，形成了特征建模方法
[2,21,22]。特征可以看作是对实体模型中几何元
素的一种归组，同组元素会被一起引用，并被赋予
特殊的设计或制造语义。总体而言，参数化建模给
实体建模主要带来以下三个益处[18,20,23]：

1. 自动的变动传播；
2. 模型/设计重用；
3. 设计、制造意图在实体模型中的表达。

参数化建模虽然有效，但模型只能在预先设计好的空间（由几何约束系统决定）里变动[24,25]。这使得参数化建模难以适用需要对模型进行自由编辑的场景，尤其是概念设计阶段[17,26-29]。针对这一问题，直接建模在 2010 年左右被提出[30]。与参数化建模中通过参数调整来间接式的修改实体模型不同，直接建模允许设计师对实体模型的几何元素进行直接式的交互编辑。直接建模方法有三大优点：

1. 直观的交互方式使其能够适用于概念设计；
2. 极高的建模自由度和效率，因为一个实体模型通过直接建模能够被变形到任意形状；
3. 高效的模型更新，因其采用局部模型更新方法。

上文简述了实体建模技术的由来与历史，下面将对实体建模、参数化建模和直接建模的关键技术进行讨论。

## 2 实体建模1.0（早期发展）

如前文所述，产品几何信息在产品计算机模型中占据着核心位置。但是在设计领域，尤其是机械设计领域，几何信息的具体内涵和定义是什么？人们在回答这一问题的过程中逐渐形成了实体建模理论与算法体系，简述如下。

### 2.1 实体数学定义

一个产品所占的空间（即实体）是三维空间 $\mathbb{R}^3$ 的一个子集，但不是所有 $\mathbb{R}^3$ 的子集都对应实体。故而需在 $\mathbb{R}^3$ 的子集上面添加约束条件，以此来剔除无实际意义的子集。这一工作主要由 Aristides A. G. Requicha 和 Herbert B. Voelcker 完成[6,9,31-33]。他们针对具有刚性、均质特性的产品提出：**实体是 $\mathbb{R}^3$ 的子集，并且具有有界性（Bounded）、半解析性（semi-analytical）和正则性（regular）。** 他们将具有这些特性的子集称为 r-set（其中"r"来自于"regular"）。

有界性是指实体的任意点到原点的距离都是有界的。例如一个 10mm×10mm×10mm 的立方体就是有界的，而 XOY 平面就是无界的。这一条件是显然的，现实世界中并不存在无限大的工业产品。

半解析性是指实体的边界由半解析曲面组成。解析曲面指曲面上每点的（某个）邻域可展开成收敛级数。这一约束条件是为了剔除如图 4 所示的高阶振荡曲面，将实体边界限定在平顺变化的曲面。半解析曲面是指曲面的边也是解析的。这是由于产品边界往往不是由一张曲面就能完整表达的，需要多张曲面缝合在一起才能表达，半解析曲面就是对缝合处的边作出具体要求。

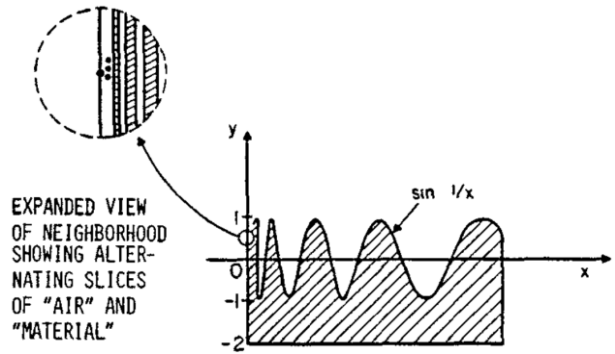

图 4 非解析曲面例子（$\sin\frac{1}{x}$）[6]

正则性是指实体是三维的，在数学上表达为实体与其内点集合的闭包是相等的，见图 5。这一约束条件是为了防止实体不包含边界点(not closed)，或者含有一维点集及二维点集，如图 6 所示。

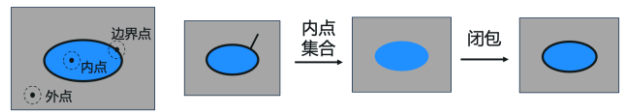

图 5 正则性定义示例

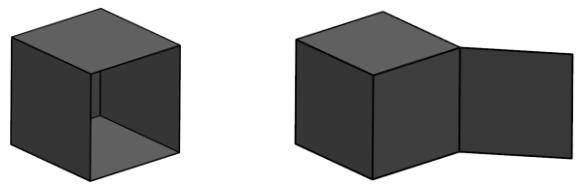

图 6 非正则实体图例

上述三个约束条件虽然可以完整刻画大部分机械产品的形状特性，但仍然允许实体具有非流形（non-manifold）边界。为此，人们又在 r-set 的基础之上添加了流形边界的约束条件，即实体边界上每个点的邻域都是二维的[34]。这在数学上表达为实体边界上每个点的（某个）邻域和二维圆盘是同胚的，而同胚指两个点集之间具有有连续的、一

一对应的映射。这一约束条件是为了防止实体出现如图 7 所示的线接触或点接触。这种情况在现实世界中是不可能的，因为其在接触处具有无穷大应力。

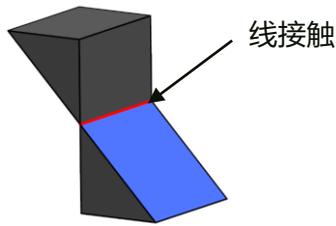

图 7  具有非流形边界的实体图例

总而言之，实体的具体内涵是一个具有有界、半解析、正则和流形性质的 $\mathbb{R}^3$ 子集，同时其所定义的对象是具有刚性、均质特性的产品。

## 2.2 实体模型表示

上述实体定义是产品几何信息的数学抽象，即数学模型，实体模型是在数学抽象基础之上的计算表示，即计算机模型（本质是一种数据结构）。在过去的 50 年里，人们提出了多种实体模型格式，详见[6,14,35,36]。其中，CSG 和 B-rep 是最常用的。

B-rep 实体模型存储实体的边界，实体的内部由边界推导而出（如使用 Winding numbers [37]、parity [38]，以及 in/out counting[36]等方法）。如图 8（a）所示，一个 B-rep 实体模型实质上仅存储组成该实体的边界面，包括这些边界面背后的几何曲面（carrying surfaces），以及这些面之间的拓扑邻接关系[10]。通过邻接关系，我们可以对曲面进行裁剪、缝合，最终生成边界面。实际的 B-rep 数据结构往往会在此基础之上添加一些冗余信息，如顶点、边以及它们间的邻接关系，以加快几何查询的速度[6]。

CSG 实体模型存储实体的构建历史，如图 8(b)所示。其使用布尔操作将多个简单实体模型组合成复杂实体模型[15]。故其主要包含两个操作：体元生成（如立方体、圆柱体）与布尔操作（如体元求交集、求并集）。与 B-rep 的显式存储方式不同，CSG 是一种隐式表示方法，模型内部只存储操作步骤，不存储操作结果，实体最终的形状需要由所记录的构建历史推导出来[39]。实际的 CSG 数据结构往往是一个二叉树，其中叶子节点存储体元的定义信息，中间节点存储布尔、刚体变换等操作。

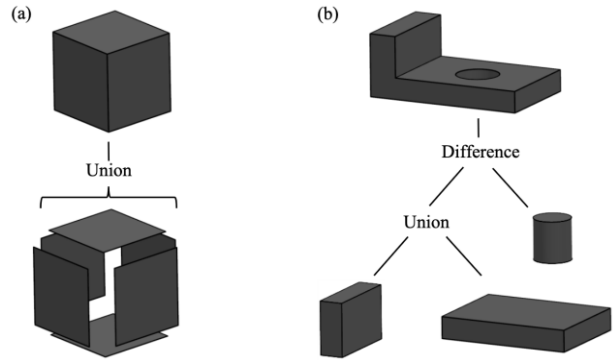

图 8  实体模型图例：（a）B-rep 和（b）CSG

总体而言，CSG 实体模型主要的优点有：
- 保证有界性质；
- 保证边界曲面半解析性质；
- 保证正则性。

其缺点有：
- 无法保证边界的流形性质，如图 9 所示；
- 模型表示不唯一（一个模型对应多个 CSG 树）。

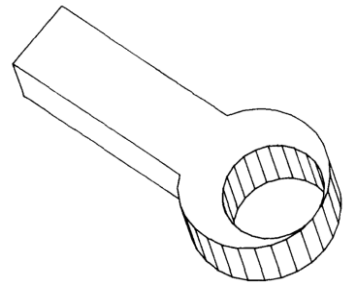

图 9  CSG 无法保证边界的流形性质[34]

B-rep 实体模型的主要优点有：
- 模型表示具有唯一性；
- （理论上）可以表示任意复杂的实体；
- 由于是显式表示，可直接用于后续模型处理。

其缺点有：
- 无法保证所存储实体模型的正则性和流形性质（一般情况下可保证有限体积性质和半解析性质）；
- 计算不鲁棒问题；
- 当模型比较复杂时，存储量比 CSG 大。

可以看出，B-rep 和 CSG 具有一定的互补性。正因为如此，现代实体建模系统一般采用 B-rep 和 CSG 相混合的表示格式[18,40 - 43]。CSG 的二叉树结构作为骨架，B-rep 模型替代了 CSG 中的体元，同时二叉树的中间节点不仅存储操作步骤，还存储部分运算结果，比如重要边界面的信息。存储这些中间

信息目的是为了将操作从布尔扩大至一些局部操作，如偏移、圆角等[16,40,44,45]。这些操作需引用被操作对象（即实体模型的某一局部几何元素），故而对这些对象进行存储或标记至关重要。

## 2.3 实体建模算法

实体建模算法指构建和编辑实体模型的算法，一般分为三个层次来实现：（1）底层数值算法；（2）中层几何/拓扑计算；（3）上层实体操作。其中上层实体操作主要包括布尔、过渡（圆角/倒角）、偏移、抽壳、扫掠、拔模、修复等[2]。这些操作在背后调用中层的几何计算或拓扑判定，主要包括求交、投影、成员判别、排序、曲面拟合等。例如，两个 B-rep 实体模型间的布尔操作实质上调用的是曲面求交和成员判别这两个操作。而几何计算/拓扑判定又会调用底层的数值算法来做解算，主要包括线性/非线性方程组求解、数值优化等通用数值算法。

此处不对具体的数值算法，几何计算/拓扑判定或者实体操作的研究现状进行详述（有兴趣的读者请见各相关方向的综述），而是对其中具有一般性的鲁棒性问题进行讨论。实体建模中的鲁棒性问题主要有三个来源：

1. 由于底层数值算法存在表示误差（来源于浮点舍入误差）、数值计算误差（来源于数值求解或优化方法，常常伴随计算步骤的增加而累加）以及中层几何计算存在不完全表达误差（来源于利用低阶曲线曲面对高阶曲线曲面进行近似所带来的误差）等表示和计算误差，以其计算结果为基础所做的逻辑性的拓扑判定可能发生不一致的情况[46-48]。这方面典型的情况有布尔操作时曲面求交误差所引起的成员判别失误。
2. 即使不存在任何表示和计算误差，拓扑判定结果也可能与几何数据不一致，并最终导致无效的实体模型。这种情况的根本原因在于 B-rep 数据结构要求拓扑和几何保持一致，才能保证模型的有效性。然而，拓扑和几何数据在 B-rep 中又是分离的，几何数据的变动不会自动反映到拓扑数据中，反之亦然[30]。正因为此，有些拓扑数据虽然从自身来看是有效的，但和几何一结合就会产生失效模型[49]。这方面典型的情况有模型修复时拓扑修正决策引发如图 7 所示模型自交（注意不是曲面自交）。
3. 永久命名问题（此处不讨论，详见 3.1 节）。

从上述讨论可以看出，实体建模中的鲁棒性问题本质是几何-拓扑不一致问题，其原因不全在数值误差，反而更在于拓扑判定的正确性。

为解决由数值误差引起的鲁棒性问题（即来源 1），一个自然的思路是将底层算法换用精确计算，如符号计算方法、有理数方法等[50,51]，但是这些方法往往在通用性或效率方面存在问题，并不实用。另一个思路是使用容差来保证即使存在计算误差，拓扑判定仍是正确的[52-54]，如图 10 所示。这种方法在理论上可完美解决几何-拓扑不一致问题，实现鲁棒建模，也是工业界所采用的方法，但是目前的容差设计方法主要以人工规则和阈值试错的方式给出，尚缺少系统的方法。特别地，容差会在多个不同局部累积增大，当它们相遇时，会发生不一致情况，进而导致错误的拓扑判定。

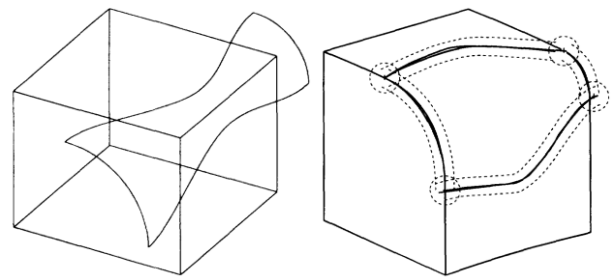

图 10 对立方体进行裁剪（左），具有容差的相交线（右）[52]

总体而言，实体建模 1.0（至 1980 年代末）奠定了实体建模的理论和算法基础。但是基础算法方面仍存在种种问题，尤其是鲁棒性问题，例如鲁棒求交、鲁棒圆角等。

---

[2] Euler 操作[16,146]也是实体建模的重要操作，但似乎现代几何建模内核里面的几何操作都不再基于 Euler 操作来实现了。

# 3 实体建模2.0（中期发展）

实体建模在 1980 年代末和 1990 年代迎来重大发展，走向参数化建模[2]。简单来说，参数化模型是在前述 B-rep 与 CSG 混合模型的基础之上增加了几何约束[20,55]。（几何约束其实早在第一代 CAD 系统"Sketchpad"上就已使用[8]。）尽管人们尝试了多种参数化建模技术，详见[18]，主流的方法由以下三部分组成[17,56,57]：

1. **2D 草图绘制**。用户首先在绘图平面上指定几何图元（点与边）的拓扑，然后在他们之间添加几何约束[58]。
2. **3D 特征生成**。对所绘制的二维草图进行拉伸、旋转等操作，以生成三维实体特征（存储为 B-rep 模型），类似 CSG 中的体元。
3. **特征组合**。将生成的 3D 特征与之前的特征进行布尔操作等，和 CSG 类似。

所有特征生成与组合的步骤又被称为建模历史，当其中一个步骤的参数发生变动，所有被记录在建模历史中的步骤都要按顺序更新，并最终生成新的实体模型。如此，我们有了参数驱动的实体模型变动。

从上述三个步骤中可以看出，参数化建模所带来的新问题主要有两个：（1）如何确保建模历史中所有引用对象的有效性，即所谓永久命名（或拓扑命名）问题；（2）如何求解用户给定的几何约束系统，即所谓几何约束求解问题。当然，还有特征识别、维护等问题，由于这些问题与高层语义更相关，而非底层的拓扑、几何、约束等，此处不展开讨论，参见[22,59,60]。

## 3.1 永久命名

在参数化建模中，当一个参数值发生变动，CAD 软件就会根据建模历史重新生成模型。由于特征间具有"父-子"依赖关系，如果这种关系在重生成过程中发生丢失或产生歧义，那么模型重生成就会失败，如图 11 中所示。这一问题在 1990 年代初被发现[24,41,61]，随后人们提出了多种方法[62,63]，但总体而言，这方面的研究很少，目前尚无系统的解决方案。

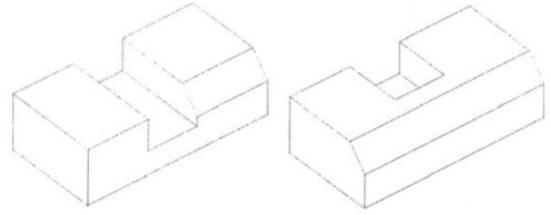

**图 11** 永久命名问题示例：参数修改前模型（右），参数修改后存在引用歧义（左）[18]

**所有永久命名方法均致力于寻找 B-rep 模型在参数变动下的不变量**。一种典型的不变量以几何元素间的拓扑关系为主，辅以他们的形状类型或相对位置，这方面典型的工作有[64-66]。后来，人们又在此基础上添加建模历史，即几何元素的生成与变动历史，提高命名的鲁棒性，这方面典型的工作有[66-72]。这一方法目前已经被工业界广泛使用。值得一提的是，其中工作[69]来自于我国华中科技大学，已成为永久命名方面的经典算法。

除上述方法外，Shapiro 将拓扑学引入永久命名的研究中，提出了拓扑不变[3]的必要条件[41,73-75]，这为研究各种永久命名方法的适用范围提供了理论基础。另外，人们还从特征语义的角度研究了参数变动下的不变量[76,77]，从限定参数变动域的角度来避免出现永久命名问题[78-80]。然而，这些工作多以纯理论研究为主，尚未见可实际应用的可能性。

混合几何、拓扑和建模历史来处理永久命名问题是目前工业界常用的方法，比如几何内核 OpenCasCade 即采用类似[64]中的方法。然而，这类方法往往需要混杂 ad hoc 规则，不够系统，也不能完整解决永久命名问题，尤其是难以处理发生大拓扑变动的情形。永久命名问题亟需新的思路，形成一个系统的解决方案。

## 3.2 几何约束求解

几何约束求解涉及两大问题：欠、过约束系统处理和恰定约束系统分解。其中约束处理是为了将用户输入的一个非恰定的约束系统修正为恰定约束系统，而约束分解是为了将一个大的约束系统分解为多个子系统，然后分别解算，以提高求解效率。

**几何约束分析与分解**。在过去的 30 余年里，人们提出了多种方法来分析几何约束系统的约束

---

[3] 实质上，Shapiro 给出的条件是允许拓扑变化的，但是这种变化需满足模型的边界面发生连续变形。

状态以及对其进行分解[43,81,82]。它们大致可以分为四类：直接求解法，逻辑推演法，图匹配法，扰动法。其中，直接求解法最为简单，利用数值计算方法（如 Newton-Raphson、homotopy）或者符号计算方法（如 Grobner bases、Wu-Ritt triangulation）对几何系统进行直接解算。如果求解成功，则为恰定约束；如果失败，则其约束状态由求解中间过程信息给出。目前，此类方法由于计算效率太低已经很少被实际采用[43]。

逻辑推演法[83,84]以一组几何公理和推演规则为基础，测试一个给定的几何约束系统是否可以被逻辑推演出来。如果成功，那么该系统是恰定的；如果存在额外的几何约束，那么该系统是过约束；反之，该系统为欠约束。这种方法本质上是将数学中的公理化思想应用到几何约束分析与求解中，具有很高的数学价值。然而，目前所制定的几何公理和推演规则离能够实际应用还远远不够。

与上述直接处理几何约束的方法不同，图匹配法将一个给定的几何约束系统首先转化为一个图，然后以图上的性质来间接的反映原几何约束系统的性质。这一方法有两大发展脉络。第一条脉络致力于在图中识别出一些特殊子图，这些子图会对应固定的几何形状或约束求解策略。这一思想首先由 Owen 提出[85]，随后[86-88]对子图种类进行了有效扩充。第二条脉络致力于使用自由度分析来提取图中恰定的子系统。这一思想首先由 Bardord[89]和 Serrano[90]提出，随后[91-93]对具体的提取算法做了补充。Hoffmann 等人在 2001 年对这两大脉络进行了详细的总结，并使用"分解-组合"这样一个抽象框架来统一表述上述方法[81]。这之后，图匹配法虽然仍有所发展，比如[94]，但整个基础框架保持不变。值得一提的是，我国中科院和华中科技大学的学者在这类方法上也做出了重要贡献，如[95,96]。

尽管图匹配法在工业界得到了广泛应用（如 DCM 和 LGS），这一方法存在重大缺陷：不能处理具有约束依赖（除了最简单的结构性依赖）的系统[97]。其原因在于当几何约束系统转化成图后，只有约束系统内的组合类信息（combinatorial information）被保留，所有几何信息均被丢弃，而很多约束依赖却与几何约束系统当时所处的几何形状息息相关。

为克服上述缺陷，扰动法于 2006 年被提出[97]。这一方法的基本思路是对约束系统的变量施加一个微小扰动，然后分析约束系统的反应，不同的反应模式就对应了不同的约束状态。扰动法最重要的结论是[97]：由于几何约束系统是非线性的，其反应随扰动施加位置的不同而有所变化，但是在一些代表性位置，扰动法的分析结果具有一般性。[98]给出了计算代表性位置的算法。这一方法由 Michelucci 等人（他也是著名的二分图匹配法和 Homotopy 法的提出者）首先提出[97,99,100]，最近在[101,102]中得到实际应用，在[103]中提升了其通用性（该问题讨论详见[104]）。完整解决其通用性问题尚需新的发展。

总体而言，几何约束分析与分解虽然在算法和应用上取得了长足进步，但几个根本问题一直未得到解决：

1. 仍缺少有效的恰定约束状态判定准则。目前广泛使用的基于自由度的判定准则缺乏理论保证，尤其是针对 3D 几何约束系统。实际例子也已经多次证实此类准则会失效[97,105]。

2. 仍无法做到最优分解。整个约束系统求解的效率由最大子系统的规模决定，因而我们需要将每个子系统的规模降到最低，但目前尚缺少有效的算法。

3. 仍无法高效求解大型 3D 几何约束系统。其原因一方面是因为缺少有效的判定准则，约束分解的鲁棒性问题突出；另一方面，传统基于分解的思路难以应对大型系统，也许并行计算是一个突破口。

4. 仍无法自动处理多解选择问题。理论上，恰定几何约束系统解的个数与约束数量是指数关系。如何自动在这么多解中选择出用户想要的解是长久以来一直存在的一个问题，这方面研究很少，而工业界多采用基于规则的方法，尚缺少系统的解决方案。

**几何约束系统处理。** 约束处理的核心任务是在系统中添加或删除约束，以消除欠、过约束状态，形成恰定约束系统，其难点在于：能够满足条件的约束往往不唯一，需要对候选约束进行设计语义方面的评价并依此作出排序。

与约束分析和分解相比，约束处理方面的研究

工作较少，进展也很小[82,106]。初期的典型工作有[92,107]，它们的思路是应用[108]中的最大加权方法，即给每个候选约束赋予一定的权重，然后从中选择那些能够形成最大加权总和的约束子集。这一方法的有效性严重依赖于权重的设计，而现有工作多采用基于 ad hoc 规则的方法，很难具有通用性。也有一些方法[109－111]是基于前述图匹配法来对约束进行选择，这些方法显然会继承图匹配法的固有缺陷。还有一些方法[112－115]是基于纯人工规则，例如约束类型等，这些策略也使这些方法缺乏通用性。最近的方法[101,102,116]是基于扰动法来做选择，但目前的进展还局限于简单约束系统[101]，或者纯过约束系统[102]，或者一般约束系统但不能完全自动化[116]。

总体而言，目前尚缺少有效的智能约束处理方法，现有方法仍处于初级发展阶段。近来人工智能的快速发展也许会给这一领域带来新的进步。

# 4 实体建模3.0（近期发展）

直接建模技术是实体建模继参数化建模后又一重要进展。这一方法虽然于 2010 年左右由工业界正式提出，但其相关的技术可追溯到 1980 年代由学术界提出的局部操作概念[16,25,45,117,118]。所谓局部操作是指对 B-rep 模型的局部几何元素直接进行编辑的方法，例如圆角、偏移等。直接建模技术就是在 tweaking 这一局部操作的基础之上发展起来的。Tweaking 允许用户对 B-rep 模型的边界面进行旋转、平移等修改，但边界面的变动被严格控制在不破坏原有拓扑关系的范围内。

直接建模放松了上述限制，以获得对模型进行任意编辑的能力（如图 12 所示），并将之重命名为 push-pull 操作。Push-pull 操作是初期直接建模技术唯一支持的操作，但目前已得到了极大的扩充，例如删除面操作等。从最近的发展来看，任何允许用户对实体模型的几何元素（点、边、面）进行直接编辑（包括移动、删除、合并、分割等）的操作都可以归到直接建模[30]。

与参数化建模中用户通过几何约束来显式、完整地表达设计意图，而计算机机械式地求解几何约束来更新模型不同，在直接建模中，用户只指定部分几何元素的目标位置（即只表达部分设计意图），其它几何元素如何协调地进行变动（主要是拓扑更新[30]）由计算机自主推断（即计算机补全用户设计意图）。如此，用户得以减负，并获得直观、快捷的模型编辑能力，而计算机是加负，需要变得更智能。如图 13 所示，如果计算机没有自主性，我们将得到一个失效模型。

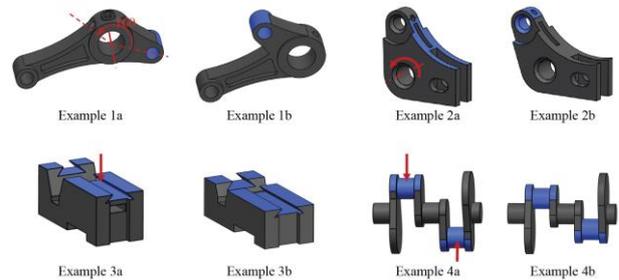

图 12 直接建模图例：（a）原模型；（b）编辑后模型[30]

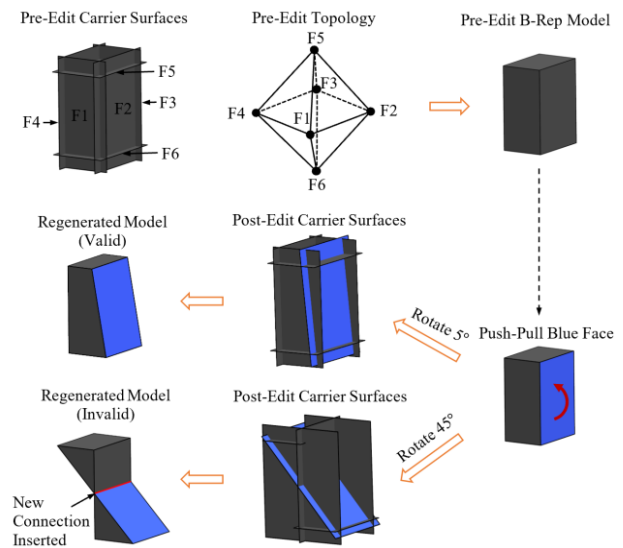

图 13 直接建模中的协调更新问题[119]

在直接建模中，关键难点在于存在多个模型更新方案，有些不能给出有效的实体模型，有些虽然可以给出有效模型，但是不符合预期（与用户设计意图不一致），一般而言，仅有一个方案是既能给出有效模型，又能与设计意图保持一致的，如图 14 所示。故而，直接建模的核心问题是方案决策问题（而参数化建模的核心问题是求解计算问题）。

针对这一问题，目前有两大思路：基于规则的以及基于连续性原理的。基于规则的方法[25,120－122]一般针对特定种类的实体模型和直接建模操作设计一组规则，来对模型进行更新。目前，这类方法已经可以鲁棒处理由平面组成的 B-rep 模型，但其它种类模型仍存在问题。基于连续性原理的方法[30,119]以模型更新时，其变动必须连续作为总要求，并将之转化成对几何元素的定量

约束条件，从而实现对模型更新方案的系统决策。目前，这类方法已经可以鲁棒处理由平面和二次曲面组成的B-rep模型，但是尚不能处理含有自由曲面的B-rep模型。

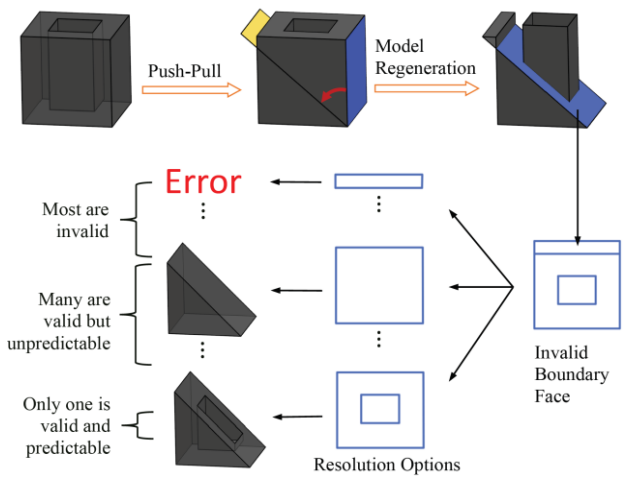

图14 直接建模中的决策问题

总体而言，直接建模最核心的问题是对模型更新方案进行决策，其关键挑战在于决策方法的系统性，从而实现鲁棒的模型更新。当前的问题主要集中于[116,123]：

1. 如何在直接编辑中保持设计语义（如边界面连接处的连续性）；
2. 如何鲁棒地直接编辑带自由曲面的实体模型；
3. 如何直接编辑参数化模型。

# 4 实体建模4.0（未来发展）

实体建模的未来发展是多样的，此处仅能就某几个方面做出讨论，分别从新方法解决旧问题、新需求带来新问题、新技术带来新发展三个方面展开。

## 4.1 从支持详细设计到支持全过程设计——参数/直接融合建模

实体建模自诞生以来就一直被诟病为不能支撑全过程设计，仅对详细设计阶段有效，对概念设计阶段帮助不大[7,124,125]。人们往往只有在已经确定了设计细节后，才使用参数化CAD软件来建模[17,26-29]。直接建模带来了直观、任意的实体模型编辑能力，使其能够支撑概念设计。

然而，在目前的CAD系统中，参数化建模功能和直接建模功能是割裂的，分别支持详细设计阶段和概念设计阶段。如何无缝融合直接建模与参数化建模，解决长期以来大家期望的在一个统一的建模方法中同时支持概念设计和详细设计，是下一代CAD建模技术亟需解决的问题。

当前的融合方法尚无法达到无缝融合[126]。最常用的方法是以参数化建模为主干，将直接建模简单地当成一个特征加入到建模历史中，如图15（a）所示。这种基于"伪特征"的融合方法，会导致原设计语义的错乱或丢失，如图15（b）所示。理想的融合方法是基于直接建模操作重定义特征模型，实现模型中设计语义的智能维护，如图15（c）所示，如此，详细设计阶段的参数语义在概念设计阶段不会被直接建模操作所破坏。

参数/直接无缝融合的关键难点在于如何将旧的几何约束系统（代表设计语义）与直接建模作用后新的B-rep模型进行同步。人们已经对此做了一些尝试[116,123,127]，在一些特定情形下取得了很好的结果，但在智能性和自动化程度上尚有很大进步空间。

## 4.2 从Computer-Aided Design到Computer-Automated Design——复杂结构建模、设计、仿真与制造

3D打印（或增材制造）技术的快速发展使得制造具有复杂微观结构的产品成为可能，如图16所示。与传统实体建模中研究的**复杂曲面**完全不同，这种**复杂结构**对建模理论、模型表示、模型操作、自动化设计都提出了新的需求，也带来了新的问题，将会给实体建模的基础理论和算法带来本质变化，主要包括：

1. 实体的定义需要扩展。如前所述，传统实体概念所定义的对象是具有刚性、均质特性的产品。而复杂结构是异质，甚至是软体或多模态的。
2. 复杂结构表示方法需要发展。复杂结构具有很高的"表面体积比"，这使得传统B-rep表示方法难以适用，否则将造成存储空间失控。如何实现复杂结构的轻量化表示，是当前复杂结构建模亟需解决的问题。目前面向复杂结构的隐式表示[128]、参数表示[129]、图表示[130]、

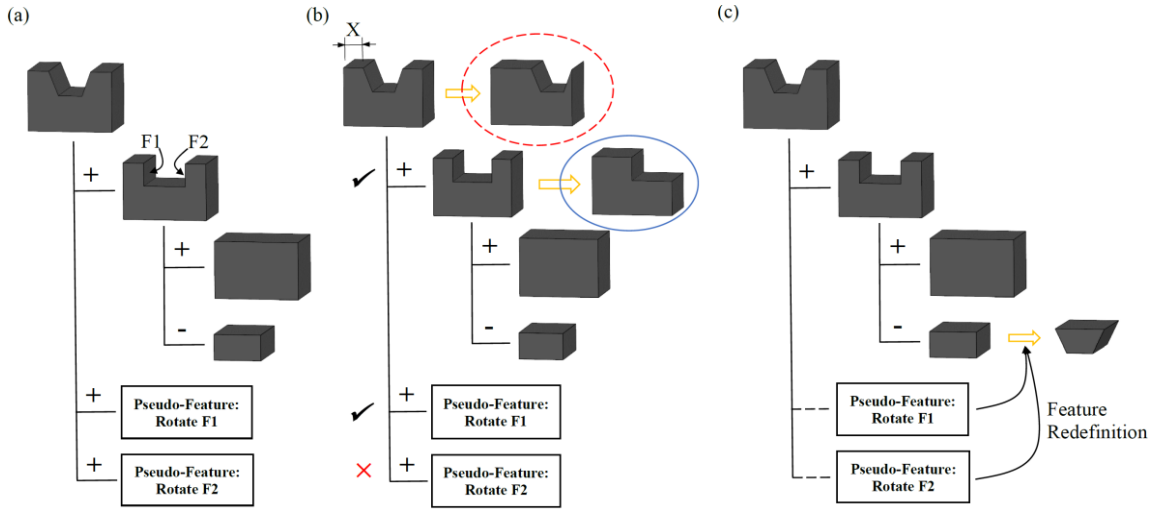

图 15  基于"伪特征"融合方法的缺陷：（a）建模历史；（b）增大 X 的尺寸不会给出红圈中理想的模型更新，而是造成模型重生成失败，其原因在于边界面 F2 的丢失（见蓝圈中模型）；（c）无缝融合应该基于直接建模操作重定义被操作特征，实现参数模型中设计语义的智能维护

网格压缩表示[131]以及它们的混合表示等方法都具有很大的局限性，往往只能处理特定类型的复杂结构[132]。

3. 复杂结构自动设计方法需要发展。复杂结构的几何、性能、工艺具有强耦合的特点，这使得其难以像传统零件一样由人来设计[30]和规划工艺[133]，而是需要由计算机来自动设计，即从 Computer-Aided Design 到 Computer-Automated Design。自动设计的难点落在 CAD/CAE/CAM 的一体化，以自动探索设计空间，达到最优结构。其要求 CAD 模型不再仅仅局限于几何模型，还要包括物理和工艺模型（如加工轨迹）。这方面的工作近年来进展很大，主要集中于结构优化[134]，但是在 CAD/CAE/CAM 的一体化方面仍有很多问题未解决，例如统一模型表示[135]、自动性能仿真[136]、工艺自动规划[137,138]等。

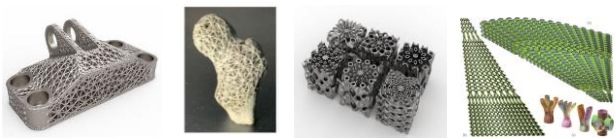

图 16  复杂结构实例[128,139–141]

### 4.3  智能 CAD、云 CAD

智能计算、云计算、并行计算以及虚拟现实等新一代计算技术将对 CAD 的发展有极大促进作用。

智能计算技术将推动 CAD 从设计师主导的人工建模向计算机主导的智能建模发展。通过机器学习，计算机将能够在一定程度上对设计师的设计、制造意图进行预测，从而实现对设计、制造语义的自动补全和识别，并将之转换为实体建模操作，这将极大地降低设计师的建模负担，尤其是在概念设计和工艺规划阶段。（本文中智能 CAD 的涵义并不包括下述功能：根据给定功能要求自动优化出产品形状或结构，这主要涉及机械式的优化算法，而非设计、制造意图的理解。）

虚拟现实技术、自然语言处理技术、计算机视觉技术将促使 CAD 系统从二维交互向三维智能交互发展，比如通过手势、语言、草图来交互。智能建模与智能交互相结合，将极大提高产品设计的效率。然而值得一提的是，这一发展不要求实体建模基础理论和算法发生本质变化，更多的是在现有 CAD 技术基础上添加一层智能技术。

云计算带来了计算资源平台的变迁，这将使得 CAD 向"计算在云上而交互在本地"的模式发展。这一发展主要会带来两个需求：（1）快速"云–端"传输；（2）并行实体建模算法。其中，需求（1）显而易见，CAD 建模的交互是实时的，故而云上的

建模结果需要快速传输到本地，并作实时绘制。需求（2）的原因在于云 CAD 并不是简单将几何建模内核搬到云上，然后针对每个用户开一个建模线程（这种方式只是在套用云概念，和桌面 CAD 没有本质区别），而是需要对用户操作（尤其是多人协同设计情况下）背后所调用的算法和资源进行管理、优化配置，合并相同类型计算，并行化不同类型计算，这本质上是要求实体建模算法向并行化发展。可以看出，云 CAD 和智能 CAD 不同，它对实体表示格式和操作算法都提出了本质变化要求。

另外，并行计算也能给实体建模中的一些老难题提供新思路。比如，针对超大规模装配模型显示和编辑问题，并行计算可部分解决其中的效率问题；比如工程设计优化慢的问题[142]。再比如，针对实体建模鲁棒操作这个历史性难题，我们可以使用并行计算的高效率来换取曲面求交等计算的高精度，从而在一定程度上解决由计算误差带来的鲁棒性问题。传统上，我们需要在精度和效率之间平衡，进而造成不可控的计算误差。

## 5 总结

本文对实体建模的发展历史作了简要梳理，按三段来介绍：早期实体建模基础理论与算法（1950 年代末到 1980 年代末），中期参数化建模（1980 年代末到 2000 年代末），以及近期直接建模（2010 年左右至今）。同时对各发展阶段的关键算法与难题以及其研究现状进行了讨论，对尚未形成系统解决方案的难题进行了重点评述。最后选取了三个方向对实体建模的未来做出展望。

值得注意的是，文中梳理的历史发展并不全面，仅涉及主要脉络。另外，未来展望也不全面，未来实体建模的发展必定是多样化而深刻的，特别是从 Computer-Aided Design 到 Computer-Automated Design 的发展趋势。这些发展也将给学术界和工业界带来众多有意义并且有趣的科研问题。

## 致谢

## 参考文献 (References)

[1] Coons S, and R. W. Mann. Computer-aided design related to the engineering design process. M.I.T. Electronic Systems Laboratory; 1960.
[2] Shah J, Mäntylä M. Parametric and Feature-Based CAD/CAM: Concepts, Techniques, and Applications. John Wiley & Sons; 1995.
[3] Coons S. An outline of the requirements for a computer-aided design system. Proceedings of the Spring Joint Computer conference, ACM; 1963, p. 299–304.
[4] Ross DT. Computer-aided design: a statement of objectives. 1960.
[5] Requicha A a G. GEOMETRIC MODELING: A First Course 6-2. Fundamental Algorithms 1999:1–28.
[6] Requicha AAG. Representations for rigid solids: theory, methods, and systems. ACM Computing Surveys 1980;12:437–64.
[7] Voelcker HB. Modeling in the Design Process. Design and Analysis of Integrated Manufacturing Systems, National Academies Press; 1988, p. 167–99.
[8] Sutherland IE. Sketchpad a man-machine graphical communication system. Simulation 1964;2:R-3.
[9] G. Requicha A, Voelcker H. Solid Modeling: Current Status and Research Directions. IEEE Computer Graphics and Applications 1983;3:25–37.
[10] Braid IC. Geometric modelling. Advances in Computer Graphics I, Springer; 1986, p. 325–62.
[11] Cohen E, Lyche T, Riesenfeld RF. MCAD: Key historical developments. Computer Methods in Applied Mechanics and Engineering 2010;199:224–8.
[12] Voelcker HB, Requicha AAG. Geometric Modeling of Mechanical Parts and Processes. Computer 1977;10:48–57.
[13] Newman W, Braid IC. The synthesis of solids bounded by many faces. Communications of the ACM 1975;18:209–16.
[14] Shapiro V. Solid modeling. Handbook of Computer Aided Geometric Design, North-Holland; 2002, p. 473–518.
[15] Requicha AG, Voelcker HB. CONSTRUCTIVE SOLID GEOMETRY. 1977.
[16] Stroud I. Boundary Representation Modelling Techniques. Springer; 2006.
[17] Camba JD, Contero M, Company P. Parametric CAD modeling: an analysis of strategies for design reusability. Computer-Aided Design 2016;74:18–31.
[18] Shah JJ. Designing with parametric CAD: classification and comparison of construction techniques. Geometric Modelling, Springer; 2001, p. 53–68.
[19] Hoffmann CM, Joan-Arinyo R. Parametric Modeling. Handbook of Computer Aided Geometric Design 2002:519–41.
[20] Roller D. An approach to computer-aided parametric design. Computer-Aided Design 1991;23:385–91.
[21] Shah JJ, Rogers MT. Functional requirements and conceptual design of the Feature-Based Modelling System. Computer-Aided Engineering Journal 1988;5:9–15.
[22] Shah JJ, Anderson D, Kim YS, Joshi S. A discourse on geometric feature recognition from cad models. Journal of Computing and Information Science in Engineering 2001;1:41–51.
[23] Camba JD, Contero M. Assessing the impact of geometric design intent annotations on parametric model alteration

activities. Computers in Industry 2015;71:35–45.
[24] Chen X, Hoffmann CM. On editability of feature-based design. Computer-Aided Design 1995;27:905–14.
[25] Rossignac JR. Issues on feature-based editing and interrogation of solid models. Computers and Graphics 1990;14:149–72.
[26] Monedero J. Parametric design: a review and some experiences. Automation in Construction 2000;9:369–77.
[27] Andrews PTJ, Shahin TMM, Sivaloganathan S. Design reuse in a CAD environment — four case studies. Computers & Industrial Engineering 1999;37:105–9.
[28] Bettig B, Bapat V, Bharadwaj B. Limitations of parametric operators for supporting systematic design. Proceedings of the 17th International Conference on Design Theory and Methodology, 2005, p. 131–42.
[29] Polly Brown. CAD: Do Computers Aid the Design Process After All? Intersect: The Stanford Journal of Science, Technology and Society n.d.;2:52–66.
[30] Zou Q, Feng H-Y. Push-pull direct modeling of solid CAD models. Advances in Engineering Software 2019;127:59–69.
[31] Requicha AAG. Mathematical models of rigid solid objects. University of Rochester; 1977.
[32] Requicha AAG. Representation of Rigid Solid Objects n.d.
[33] Requicha AAG, Voelcker HB. Solid Modeling: A Historical Summary and Contemporary Assessment. IEEE Computer Graphics and Applications 1982;2:9–24.
[34] Mantyla M. A note on the modeling space of Euler operators. Computer Vision, Graphics, and Image Processing 1984;26:45–60.
[35] Baer A, Eastman C, Henrion M. Geometric modelling: a survey. Computer-Aided Design 1979;11:253–72.
[36] Botsch, Mario, Leif Kobbelt, Mark Pauly, Pierre Alliez and BL. Polygon mesh processing. CRC press; 2010.
[37] Jacobson A, Kavan L, Sorkine-Hornung O. Robust inside-outside segmentation using generalized winding numbers. ACM Transactions on Graphics (TOG) 2013;32.
[38] Bridson R, Fedkiw R, Anderson J. Robust Treatment of Collisions, Contact and Friction for Cloth Animation n.d.
[39] Requicha AAG, Voelcker HB. Boolean operations in solid modeling: boundary evaluation and merging algorithms. Proceedings of the IEEE 1985;73:30–44.
[40] Rossignac JR, Requicha AAG. Offsetting operations in solid modelling. Computer Aided Geometric Design 1986;3:129–48.
[41] Shapiro V, Vossler DL. What is a parametric family of solids? Proceedings of the 3rd ACM Symposium on Solid Modeling and Applications, 1995, p. 43–54.
[42] Bodein Y, Rose B, Caillaud E. Explicit reference modeling methodology in parametric CAD system. Computers in Industry 2014;65:136–47.
[43] Bettig B, Hoffmann CM. Geometric constraint solving in parametric computer-aided design. Journal of Computing and Information Science in Engineering 2011;11:021001.
[44] Braid I. Non-local blending of boundary models I C Braid. Computer-Aided Design 1997;29:89–100.
[45] Grayer AR. Alternative approaches in geometric modelling. Computer-Aided Design 1980;12:189–92.
[46] Michelucci D. An Introduction to the Robustness Issue n.d.
[47] Hoffmann CM. Robustness in Geometric Computations. Journal of Computing and Information Science in Engineering 2001;1:143–55.
[48] Rossignac JR. Through the cracks of the solid modeling milestone. From Object Modeling to Advanced Visualization 1994:1–75.
[49] Shen G, Sakkalis T, Patrikalakis N. Analysis of boundary representation model rectification. Proceedings of the 6th ACM Symposium on Solid Modeling and Applications, 2001, p. 149–58.
[50] Berberich E, Eigenwillig A, Hemmer M, Hert S, Kettner L, Mehlhorn K, et al. EXACUS: Efficient and exact algorithms for curves and surfaces. Lecture Notes in Computer Science 2005;3669:155–66.
[51] Keyser J, Culver T, Foskey M, Krishnan S, Manocha D. ESOLID—a system for exact boundary evaluation. Computer-Aided Design 2004;36:175–93.
[52] Jackson DJ. Boundary representation modelling with local tolerances. Symposium on Solid Modeling and Applications, 1995, p. 247–53.
[53] Stroud I, Nagy H. Solid Modeling and CAD Systems : How to Survive a CAD System. vol. 1. 2011.
[54] Qi J, Shapiro V. ε-Topological formulation of tolerant solid modeling. Computer-Aided Design 2006;38:367–77.
[55] Rossignac JR. Constraints in constructive solid geometry. Proceedings of the 1986 Workshop on Interactive 3D Graphics - SI3D '86 1987:93–110.
[56] Mun D, Han S, Kim J, Oh Y. A set of standard modeling commands for the history-based parametric approach. Computer-Aided Design 2003;35:1171–9.
[57] Hoffmann CM, Joan-Arinyo R. On user-defined features. Computer-Aided Design 1998;30:321–32.
[58] Bettig B, Shah J. Derivation of a standard set of geometric constraints for parametric modeling and data exchange. Computer-Aided Design 2001;33:17–33.
[59] Bidarra R, Bronsvoort WF. Semantic feature modelling. Computer-Aided Design 2000;32:201–25.
[60] Li L, Zheng Y, Yang M, Leng J, Cheng Z, Xie Y, et al. A survey of feature modeling methods: Historical evolution and new development. Robotics and Computer-Integrated Manufacturing 2020;61:101851.
[61] Hoffmann CM. On the Semantics of Generative Geometry Representations. Proceedings of the ASME Design Engineering Technical Conference 2021;Part F167972-14:411–9.
[62] Marcheix D, Pierra G. A survey of the persistent naming problem. Proceedings of the Symposium on Solid Modeling and Applications 2002:13–22.
[63] Farjana SH, Han S. Mechanisms of Persistent Identification of Topological Entities in CAD Systems: A Review. Alexandria Engineering Journal 2018;57:2837–49.
[64] Capoyleas V, Chen X, Hoffmann CM. Generic naming in generative, constraint-based design. Computer-Aided Design 1996;28:17–26.
[65] Wang Y, Nnaji BO. Geometry-based semantic ID for persistent and interoperable reference in feature-based parametric modeling. Computer-Aided Design 2005;37:1081–93.
[66] Mehdi B-A, Marcheix D, Skapin X, Baba-Ali M. A Method to Improve Matching Process by Shape Characteristics in

Parametric Systems. Computer-Aided Design and Applications 2009;6:341–50.

[67] Kripac J. A mechanism for persistently naming topological entities in history-based parametric solid models. Computer-Aided Design 1997;29:113–22.

[68] Agbodan D, Marcheix D, Pierra G, Thabaud C. A topological entity matching technique for geometric parametric models. 2003 Shape Modeling International., IEEE Comput. Soc; n.d., p. 235–44.

[69] Wu J, Zhang T, Zhang X, Zhou J. A face based mechanism for naming, recording and retrieving topological entities. Computer-Aided Design 2001;33:687–98.

[70] Mun DW, Han SH. Identification of Topological Entities and Naming Mapping for Parametric CAD Model Exchanges. International Journal of CAD/CAM 2005;5:69–81.

[71] Cheon S, Mun DW, Han SH, Kim BC. Name matching method using topology merging and splitting history for exchange of feature-based CAD models. Journal of Mechanical Science and Technology 2012;26:3201–12.

[72] Farjana SH, Han S, Mun D. Implementation of persistent identification of topological entities based on macro-parametrics approach. Journal of Computational Design and Engineering 2016;3:161–77.

[73] Raghothama S, Shapiro V. Boundary representation deformation in parametric solid modeling. ACM Transactions on Graphics 1998;17:259–86.

[74] Raghothama S, Shapiro V. Topological framework for part families. Journal of Computing and Information Science in Engineering 2002;2:246–55.

[75] Raghothama S, Shapiro V. Necessary conditions for boundary representation variance. Proceedings of the thirteenth annual symposium on Computational geometry - SCG '97, New York, New York, USA: ACM Press; 1997, p. 77–86.

[76] Bidarra R, Bronsvoort WF. Persistent naming through persistent entities. Proceedings - Geometric Modeling and Processing: Theory and Applications, GMP 2002 2002:233–40.

[77] Bidarra R, Nyirenda PJ, Bronsvoort WF. A Feature-Based Solution to the Persistent Naming Problem. CAD Solutions LLC 2013;2:517–26.

[78] Van der Meiden HA, Bronsvoort WF. Tracking topological changes in parametric models. Computer Aided Geometric Design 2010;27:281–93.

[79] Hoffmann CM, Kim KJ. Towards valid parametric CAD models. Computer-Aided Design 2001;33:81–90.

[80] Tang Z, Zou Q, Gao S. Towards computing complete parameter ranges in parametric modeling 2022.

[81] Hoffman CM, Lomonosov A, Sitharam M. Decomposition plans for geometric constraint systems, part I: performance measures for CAD. Journal of Symbolic Computation 2001;31:367–408.

[82] Zou Q, Tang Z, Feng H-Y, Gao S, Zhou C, Liu Y. A review on geometric constraint solving 2022.

[83] Dufourd J-F, Mathis P, Schreck P. Geometric construction by assembling solved subfigures. Artificial Intelligence 1998;99:73–119.

[84] Joan-Arinyo R, Soto A. A correct rule-based geometric constraint solver. Computers & Graphics 1997;21:599–609.

[85] Owen JC. Algebraic solution for geometry from dimensional constraints. Proceedings of the First ACM symposium on Solid Modeling Foundations and CAD/CAM Applications, 1991, p. 397–407.

[86] Bouma W, Fudos I, Hoffmann C, Cai J, Paige R. Geometric constraint solver. Computer-Aided Design 1995;27:487–501.

[87] Fudos I, Hoffmann CM. A graph-constructive approach to solving systems of geometric constraints. ACM Transactions on Graphics 1997;16:179–216.

[88] Gao X-S, Hoffmann CM, Yang W-Q. Solving spatial basic geometric constraint configurations with locus intersection. Proceedings of the Seventh ACM Symposium on Solid Modeling and Applications, 2002, p. 111–22.

[89] Bardord LA. A graphical, language-based editor for generic solid models represented by constraints. Cornell University, 1987.

[90] Serrano D. Constraint management in conceptual design. Massachusetts Institute of Technology, 1987.

[91] Ait-Aoudia S, Jegou R, Michelucci D. Reduction of constraint systems. Proceedings of Compugraphics, 1993, p. 83–92.

[92] Latham RS, Middleditch AE. Connectivity analysis: a tool for processing geometric constraints. Computer-Aided Design 1996;28:917–28.

[93] Hoffmann CM, Lomonosov A, Sitharam M. Finding solvable subsets of constraint graphs. Proceedings of International Conference on Principles and Practice of Constraint Programming, 1997, p. 463–77.

[94] Hidalgo M, Joan-Arinyo R. h-graphs: A new representation for tree decompositions of graphs. Computer-Aided Design 2015;67–68:38–47.

[95] Gao X-S, Lin Q, Zhang G-F. A C-tree decomposition algorithm for 2D and 3D geometric constraint solving. Computer-Aided Design 2006;38:1–13.

[96] Xia H, Wang B, Chen L, Huang Z. 3D geometric constraint solving using the method of kinematic analysis. The International Journal of Advanced Manufacturing Technology 2006 35:7 2006;35:711–22.

[97] Michelucci D, Foufou S. Geometric constraint solving: the witness configuration method. Computer Aided Design 2006;38:284–99.

[98] Kubicki A, Michelucci D, Foufou S. Witness computation for solving geometric constraint systems. Proceedings of the 2014 Science and Information Conference, 2014, p. 759–70.

[99] Thierry SEB, Schreck P, Michelucci D, Funfzig C, Génevaux JD. Extensions of the witness method to characterize under-, over- and well-constrained geometric constraint systems. Computer Aided Design 2011;43:1234–49.

[100] Foufou S, Michelucci D. Interrogating witnesses for geometric constraint solving. Information and Computation 2012;216:24–38.

[101] Moinet M, Mandil G, Serre P. Defining tools to address over-constrained geometric problems in Computer Aided Design. Computer-Aided Design 2014;48:42–52.

[102] Hu H, Kleiner M, Pernot J-P. Over-constraints detection and resolution in geometric equation systems. Computer-Aided Design 2017;90:84–94.

[103] Zou Q, Feng H-Y. Variational B-rep model analysis for direct modeling using geometric perturbation. Journal of

[104] Zou Q, Feng H-Y. On Limitations of the Witness Configuration Method for Geometric Constraint Solving in CAD Modeling 2019.

[105] Haller K, Lee-St.john A, Sitharam M, Streinu I, White N. Body-and-cad geometric constraint systems. Computational Geometry 2012;45:385–405.

[106] Hu H, Kleiner M, Pernot JP, Zhang C, Huang Y, Zhao Q, et al. Geometric Over-Constraints Detection: A Survey. Archives of Computational Methods in Engineering 2021;28:4331–55.

[107] Jermann C, Hosobe H. A constraint hierarchies approach to geometric constraints on sketches. Proceedings of the 2008 ACM Symposium on Applied Computing, ACM; 2008, p. 1843–4.

[108] Borning A, Freeman-Benson B, Wilson M. Constraint hierarchies. Lisp and Symbolic Computation 1992;5:223–70.

[109] Joan Arinyo R, Soto Riera A, Vila Marta S, Vilaplana Pasto J. Transforming an under-constrained geometric constraint problem into a well-constrained one. Proceedings of the eighth ACM Symposium on Solid Modeling and Applications, ACM; 2002, p. 33–44.

[110] Hoffmann CM, Sitharam M, Yuan B. Making constraint solvers more usable: overconstraint problem. Computer-Aided Design 2004;36:377–99.

[111] Zhang G-F, Gao X-S. Well-constrained Completion and Decomposition for under-constrained Geometric Constraint Problems. International Journal of Computational Geometry & Applications 2006;16:461–78.

[112] Murugappan S, Sellamani S, Ramani K. Towards beautification of freehand sketches using suggestions. Proceedings of the 6th Eurographics Symposium on Sketch-Based Interfaces and Modeling, ACM; 2009, p. 69–76.

[113] Mills BI, Langbein FC, Marshall AD, Martin RR. Estimate of frequencies of geometric regularities for use in reverse engineering of simple mechanical components. 2001.

[114] Zou HL, Lee YT. Constraint-based beautification and dimensioning of 3D polyhedral models reconstructed from 2D sketches. Computer-Aided Design 2007;39:1025–36.

[115] Langbein FC, Marshall AD, Martin RR. Choosing consistent constraints for beautification of reverse engineered geometric models. Computer-Aided Design 2004;36:261–78.

[116] Zou Q, Feng H-Y. A decision-support method for information inconsistency resolution in direct modeling of CAD models. Advanced Engineering Informatics 2020;44:101087.

[117] Stroud I, Xirouchakis PC. CAGD - Computer-aided gravestone design. Advances in Engineering Software 2006;37:277–86.

[118] Fahlbusch K-P, Roser TD. HP PE/SolidDesigner: dynamic modeling for three-dimensional computer-aided design. Hewlett-Packard Journal 1995;46:6–13.

[119] Zou Q, Feng H-Y. A robust direct modeling method for quadric B-rep models based on geometry-topology inconsistency tracking. Engineering with Computers 2021:1–14.

[120] Woo Y, Lee SH. Volumetric modification of solid CAD models independent of design features. Advances in Engineering Software 2006;37:826–35.

[121] Kim BC, Mun DW. Stepwise volume decomposition for the modification of B-rep models. The International Journal of Advanced Manufacturing Technology 2014;75:1393–403.

[122] Fu J, Chen X, Gao S. Automatic synchronization of a feature model with direct editing based on cellular model. Computer-Aided Design and Applications 2017;14:680–92.

[123] Qin X, Tang Z, Gao S. Automatic update of feature model after direct modeling operation. Computer-Aided Design and Applications 2021;18:170–85.

[124] Zou Q, Zheng Q, Tang Z, Gao S. Variational design for a structural family of CAD models 2022.

[125] Chung JCH, Hwang T-S, Wu C-T, Jiang Y, Wang J-Y, Bai Y, et al. Framework for integrated mechanical design automation. Computer-Aided Design 2000;32:355–65.

[126] Zou Q. Parametric/direct CAD integration. ArXiv 2022.

[127] Zou Q. VARIATIONAL DIRECT MODELING FOR COMPUTER-AIDED DESIGN. The University of British Columbia, 2019.

[128] Ding J, Zou Q, Qu S, Bartolo P, Song X, Wang CCL. STL-free design and manufacturing paradigm for high-precision powder bed fusion. CIRP Annals 2021;70:167–70.

[129] Massarwi F, Machchhar J, Antolin P, Elber G. Hierarchical, random and bifurcation tiling with heterogeneity in micro-structures construction via functional composition. Computer-Aided Design 2018;102:148–59.

[130] Gupta A, Allen G, Rossignac J. QUADOR: QUADric-Of-Revolution beams for lattices. Computer-Aided Design 2018;102:160–70.

[131] Chougrani L, Pernot JP, Véron P, Abed S. Lattice structure lightweight triangulation for additive manufacturing. Computer-Aided Design 2017;90:95–104.

[132] Dong G, Tang Y, Zhao YF. A survey of modeling of lattice structures fabricated by additive manufacturing. Journal of Mechanical Design, Transactions of the ASME 2017;139.

[133] Su C, Jiang X, Huo G, Zou Q, Zheng Z, Feng HY. Accurate model construction of deformed aero-engine blades for remanufacturing. International Journal of Advanced Manufacturing Technology 2020;106:3239–51.

[134] Wu J, Sigmund O, Groen JP. Topology optimization of multi-scale structures: a review. Structural and Multidisciplinary Optimization 2021;63:1455–80.

[135] Hughes TJR, Cottrell JA, Bazilevs Y. Isogeometric analysis: CAD, finite elements, NURBS, exact geometry and mesh refinement. Computer Methods in Applied Mechanics and Engineering 2005;194:4135–95.

[136] Shapiro V, Tsukanov I, Grishin A. Geometric Issues in Computer Aided Design/Computer Aided Engineering Integration. Journal of Computing and Information Science in Engineering 2011;11:021005.

[137] Zhao J, Zou Q, Li L, Zhou B. Tool path planning based on conformal parameterization for meshes. Chinese Journal of Aeronautics 2015;28:1555–63.

[138] Zou Q. Length-optimal tool path planning for freeform surfaces with preferred feed directions based on Poisson formulation. Computer-Aided Design 2021;139:103072.

[139] Liu S, Liu T, Zou Q, Wang W, Doubrovski EL, Wang CCL. Memory-Efficient Modeling and Slicing of Large-Scale Adaptive Lattice Structures. Journal of Computing and Information Science in Engineering 2021;21.

[140] Gupta A, Allen G, Rossignac J. QUADOR: QUADric-Of-Revolution beams for lattices. Computer-Aided Design 2018;102:160–70.

[141] Massarwi F, Machchhar J, Antolin P, Elber G. Hierarchical, random and bifurcation tiling with heterogeneity in micro-structures construction via functional composition. Computer-Aided Design 2018;102:148–59.

[142] Wang G, Zou Q, Zhao C, Liu Y, Ye X. A Highly Efficient Approach for Bi-Level Programming Problems Based on Dominance Determination. Journal of Computing and Information Science in Engineering 2022;22.

[143] Zou Q, Zhang J, Deng B, Zhao J. Iso-level tool path planning for free-form surfaces. Computer-Aided Design 2014;53:117–25.

[144] Zou Q, Zhao J. Iso-parametric tool-path planning for point clouds. Computer-Aided Design 2013;45:1459–68.

[145] Zou Q. Robust and efficient tool path generation for machining low-quality triangular mesh surfaces. International Journal of Production Research 2020;59:7457–67.

[146] Eastman CM, Weiler KJ. Geometric modeling using the Euler operators. Carnegie Mellon University; 1979.